\documentclass{ws-procs975x65}
\usepackage{amsmath}
\def\beq{\begin{equation}}
\def\eeq{\end{equation}}

\begin{document}

\title{The future Milky Way and Andromeda galaxy merger}
\author{Riccardo Schiavi$^*$ and Roberto Capuzzo Dolcetta}

\address{Department of Physics, Universit\`a La Sapienza,\\
Roma, 00185, Italy\\
$^*$E-mail: riccardo.schiavi@uniroma1.it}

\author{Manuel Arca Sedda} 

\address{ARI, University of Heidelberg,\\
Heidelberg, 69117, Germany}

\begin{abstract}
According to our current knowledge about physical and dynamical properties of the Milky Way-M31 system, it seems likely that these two galaxies will collide and eventually merge in a time very sensitive to initial conditions. Using the HiGPUs code\cite{Capuzzo}, we have performed several numerical simulations to study the dynamics of the system, trying to define the role of indeterminacy in the present day observed relative velocities of the two galaxies and the time of the merger. At the same time, we have followed the dynamics of the two massive black holes sitting in the galactic centers, to check (within the space and time resolution limits of our simulation) their relative motion upon the completion of the galaxies merger process.
\end{abstract}

\keywords{galaxy merger; supermassive black holes; Milky Way-Andromeda collision}

\bodymatter


\section{Purposes of the work}

In this work we aim to follow the evolution of the ``binary'' system composed by our own galaxy, the Milky Way (MW), and the nearby Andromeda galaxy (M31). Observational data are currently not sufficient to strictly constrain the dynamics of this interaction and, therefore, a wide range of scenarios for their future evolution are possible. In any case, almost all of them imply an eventual merger between the two galaxies in a time which is very dependent upon parameters of the simulation. 

 In our study we  considered a set of initial conditions spanning a range of values of the tangential component of the relative velocity of the MW and M31. This allowed us to quantify how these differences affect the interaction.
A particular subtopic of relevant interest is the motion of the two massive black holes
(SMBHs)   known to be present in the two galaxies central regions, with especial regard to the last phase of the galactic merging process, when the two  SMBHs are expected to  form a close binary system. The results we obtained so far for the evolution of their orbits during the galaxies interaction, are interesting but the present simulation resolution is not enough to represent the very final stage, down to what is called the ``final parsec problem''.



\section{Methods}

For these simulations we have used the HiGPUs code\cite{Capuzzo}, which  performs direct $N$-body calculations exploiting a full parallelization on hybrid platforms  (CPUs and Graphic Processing Units). So far, we have used a number of particles of order  $3.2\times10^4$, aiming to extend resolution by reaching, at least, one million of particles. Initial conditions are obtained with the code ``GSAM'' (Galaxy SAMpler), written by Arca-Sedda et al. (2015), and able to generate star clusters models with different density distributions. In our case the two galaxies are modelled according to a King density profile.

We have performed three different simulations at varying the transverse velocity of Andromeda as seen by the MW,  keeping unchanged the other parameters. 

In Table 1 all the initial parameters, common to the three simulations, are shown.

\begin{table}[h]
\centering
\caption{table}{Table 1}\\
\smallskip
\begin{tabular}{l|cc}
      &\textbf{Milky Way}  & \textbf{Andromeda}\\ 
\hline
Mass ($M_\odot$)          &$9.0\times10^{11}$             &$1.5\times10^{12}$\\
$W_0$       &6        &6\\
King radius $r_c$ (kpc)            &1.535      &3.385\\

Relative radial velocity (km/s)    &\multicolumn{2}{c}{-120}\\
Initial separation (kpc)    &\multicolumn{2}{c}{780}\\
\hline
\end{tabular}
\bigskip
\caption{table}{Common initial conditions in the three simulations}
\label{tab:properties}
\end{table}

\section{Preliminary results}

Modelling the two galaxies  orbital evolution, we could investigate how the merger time changes at varying the relative initial transverse velocity ($V_{t0}$), as summarized in Table 2. Our results show that the great uncertainty on this kinematical parameter makes the time of merger undetermined for a factor $\sim 80$. Actually, a wide range of merger times are found in the literature: in some cases \cite{vdm2012} a very low transverse velocity was reported ($V_{t0}\approx17 km/s$), whereas in others \cite{Sal2016} a value about 10 times larger,  $V_{t0}\approx164 km/s$. According to the most recent estimations \cite{vdm2019} a reliable value seems  $V_{t0}\approx57 km/s$, but the level of uncertainty is still very large. For these reasons, our results are only indicative, as upper and lower limits. 
Moreover, in these present simulations the intergalactic medium (IGM) has not been considered, which, on its side, should act to reduce the time for the merger of the two galaxies.

\begin{table}[h]
\centering
\caption{table}{Table 2}\\
\smallskip
\begin{tabular}{c|c}
\toprule
$\mathbf{V_{\mathbf{t0}}}$ \textbf{(km/s)}  & \textbf{Time of the merger (Gyr)}\\ 
\hline
20          &6.7 \\
50      &108.8\\
80       &544.6\\
\hline
\end{tabular}
\bigskip
\caption{table}{Time of the merger for the three initial tangential velocities of M31.}
\label{tab:time}
\end{table}

\begin{figure}
\centering
\includegraphics[scale=0.5]{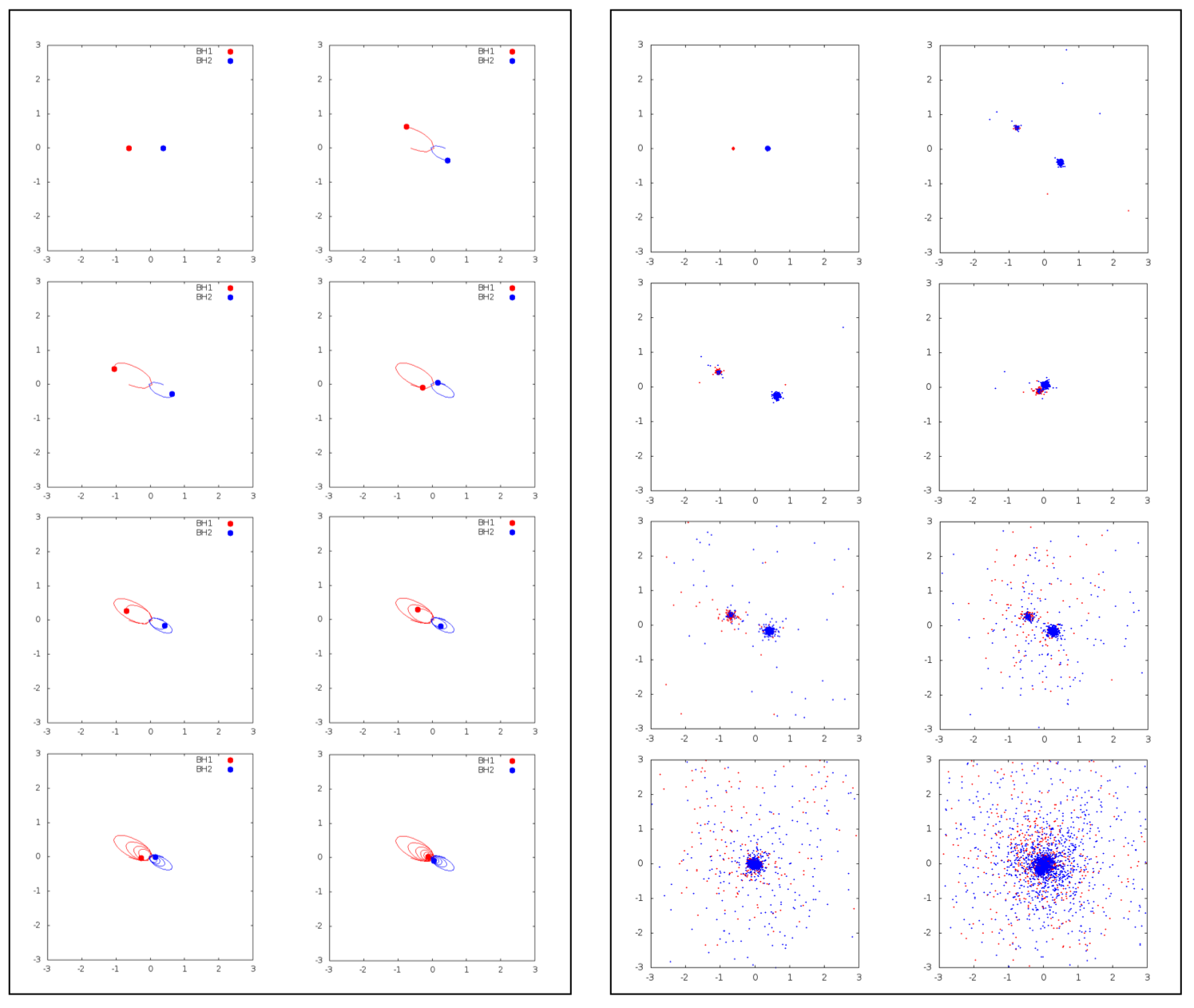}
\caption{Snapshots at various times of the configuration of the two BHs (left columns) and their host galaxies (right columns) in the case of $V_{t0}=50 km/s$. Red colour refers to the Milky Way, blue to Andromeda. The unit of distance is 780 kpc. The time interval between each snapshot is 12.5 Gyr.}
\label{fig:1}
\end{figure}

\begin{figure}
\centering
\includegraphics[scale=0.7]{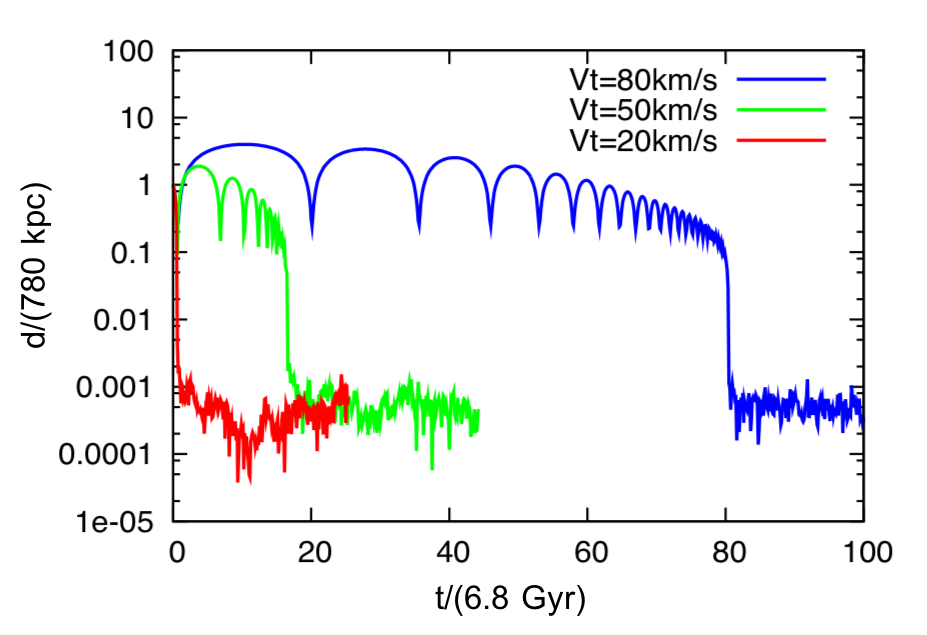}
\caption{Distance between the two BHs as function of time, for three different initial tangential velocities.}
\label{fig:2}
\end{figure}

Figure~\ref{fig:1} plots some snapshots of the merging process in the case with $V_{t0}=50$ km/s: in the left panel we show the relative motion of only the two central SMBHs while, on the right, the two interacting galaxies are shown.

We have plotted in Fig.~\ref{fig:2} the time evolution of the distance between the two SMBHs in the three cases of $V_{t0}$ examined. It should be noticed that, after the merger of their host galaxies, the two SMBHs keep orbiting around each other at almost constant distance (roughly circular orbit). The reason why the SMBHs binary does not experience a further shrink is due to the loss of efficiency of gravitational encounters, because, due to unsufficient number of simulation particles, there are too few ``stars'' enclosed within the scale the two SMBHs are orbiting. This means that to reproduce satisfactorily the inner dynamics of the SMBHs binary we need at least a factor 10 larger $N$.

Anyway, an interesting result is that the distance at which the binary stalls (around 300 pc) seems to be independent of the initial tangential velocity, because it likely depends only on $N$ and not on the other initial conditions.

\section{Tentative Conclusions}

This short note reports some of our current work on the dynamics of the future merging process of our Galaxy and Andromeda. Since the tangential velocity of Andromeda is not precisely constrained by observations, we decided to perform three simulations varying this parameter. Depending of this, we have found that the time of the merger results great. This is partly due to that in these preliminary simulations we did not consider the role of the intergalactic medium, which, according to some previous simulations\cite{cl2008}, can speed up significantly the merger process, via dynamical friction.

The relative motion of the two central SMBHs shows a regular behaviour on a large spatial scale, but we should increase the resolution to extend reliably the study also to the innermost region of the new cluster formed after the merger.
Actually, we are planning other simulations with larger values of $N$,  extending the range of values for $V_{t0}$, and with the inclusion of the intergalactic medium dynamical friction effect.

\end{document}